\documentclass[conference]{IEEEtran}
\IEEEoverridecommandlockouts
\usepackage{cite}
\usepackage{amsmath,amssymb,amsfonts}
\usepackage{graphicx}
\usepackage{textcomp}
\usepackage{xcolor}
\usepackage{algorithm}
\usepackage{algpseudocode}
\def\BibTeX{{\rm B\kern-.05em{\sc i\kern-.025em b}\kern-.08em
    T\kern-.1667em\lower.7ex\hbox{E}\kern-.125emX}}
\begin{document}

\title{Optimizing Global Quantum Communication via Satellite Constellations}

\author{
    \IEEEauthorblockN{Yichen Gao\IEEEauthorrefmark{1}, Guanqun Song\IEEEauthorrefmark{1}, Ting Zhu\IEEEauthorrefmark{1}}
    \IEEEauthorblockA{\IEEEauthorrefmark{1}\textit{Department of Computer Science and Engineering}, \textit{The Ohio State University}, Columbus, USA \\
    Email: gao.2174@osu.edu, song.2107@osu.edu, zhu.3445@osu.edu}
}

\maketitle

\begin{abstract}
In this paper, we investigate the optimization of global quantum communication through satellite constellations. We address the challenge of quantum key distribution (QKD) across vast distances and the limitations posed by terrestrial fiber-optic networks. Our research focuses on the configuration of satellite constellations to improve QKD between ground stations and the application of innovative orbital mechanics to reduce latency in quantum information transfer. We introduce a novel approach using quantum relay satellites in Molniya orbits, enhancing communication efficiency and coverage. The use of these high-eccentricity orbits allows us to extend the operational presence of satellites over targeted hemispheres, thus maximizing the quantum network's reach. Our findings provide a strategic framework for deploying quantum satellites and relay systems to achieve a robust and efficient global quantum communication network.
\end{abstract}

\section{Introduction}

Quantum Key Distribution (QKD)\cite{BEN84,shor2000simple,ren2017ground} is a cryptographic key exchange. It utilizes the principles of quantum mechanics to ensure the confidentiality of the communication channel. The key of QKD's security guarantee is the no-cloning theorem and the Heisenberg uncertainty principle. This ensures the non-replicability of quantum states with the immunity of quantum systems to jamming. This ensures the protection of information from attacks. Quantum communication is currently at the forefront of secure communication technology. In the era of advancing quantum computing power, traditional communication encryption algorithms are constantly being challenged\cite{gisin2007quantum}, and the theoretical confidentiality of quantum communication can provide unbreakable protection for important information in the future.

The BB84 protocol, developed by Bennett and Brassard in 1984, employs the polarization states of photons to securely distribute cryptographic keys\cite{BEN84}. The E91 protocol, introduced by Ekert\cite{ekert1991quantum} in 1991, harnesses the peculiarities of quantum entanglement and Bell’s inequalities to offer an alternative secure key exchange mechanism. Lucamarini et al\cite{lucamarini2018overcoming} proposed Twin-Field QKD (TF-QKD) to enhance the achievable distance of secure communication by utilizing entangled photon pairs sent to a central relay. The continuous-variable quantum key distribution (CV-QKD)\cite{weedbrook2012gaussian} protocol, employing continuous-variable quantum states, offers an alternative to the single-photon encoding approach, and might simplify the implementation and compatibility with existing communication devices.
 
 The work of Gisin et al.\cite{gisin2002quantum}, showcased the feasibility of implementing QKD over increasingly long distances using optical fibers. The communication distance of the quantum communication is extended to hundreds of kilometers\cite{tang2014measurement,tang2014measurement}. However, due to the path loss of photons through the fiber material due to impurities and inherent fiber properties (e.g., reflection and scattering). These factors exacerbate photon losses and severely limit the effective distance over which quantum communications can be reliably maintained. As a result, the reach of current fiber-based QKD systems is typically limited to global-scale communication. Therefore, the successful application of QKD in free-space channels (satellite-to-ground communication) has attracted considerable interest. The channel attenuation produced by light counting in a vacuum environment is much smaller than in an atmosphere with gas disturbances, the Micius\cite{liao2017satellite, yin2017satellite, liao2018satellite} satellite demonstrated intercontinental QKD, establishing ultra-secure communication links over thousands of kilometers, well beyond the limits of terrestrial fiber-optic networks. Based on this background, more researchers began to focus on satellite-based quantum communication networks\cite{chen2021integrated, bae2023blockwise, de2023satellite} and satellite-to-ground station downlink analysis\cite{dequal2021feasibility, panigrahy2022optimal, liao2018satellite}. In satellite quantum communication, the communication relationship between the satellite and the ground station is affected by the loss, error rate, and noise (signal-to-noise ratio) in the quantum channel. The channel attenuation in the communication process is caused by beam expansion or divergence in free space. In the downlink, most of the beam propagation occurs in the vacuum, where the beam maintains its diffraction-limited properties and encounters a turbulent atmosphere only during the last tens of kilometers of its path. In contrast, for uplinks, the wavefront is distorted at the beginning of its path, which has a stronger effect on beam propagation.Part of the current research is based on dual downlink architectures in satellite constellations\cite{khatri2021spooky, panigrahy2022optimal, dequal2021feasibility}. A satellite is used as a signal generator to distribute pairs of photons to two different ground stations, which are distributed to the ground station and generate secret keys. Khatri et al.\cite{khatri2021spooky} analyzed satellite configuration in a dual downlink architecture based on satellite constellations to improve the secret key shengcheng between a pair of ground stations.Nitish et al.\cite{panigrahy2022optimal} Dynamically assigning satellites to pairs of ground stations maximizes the overall performance of the network for a given cluster of satellites and group of ground stations. However, Nitish's work does not take into account the geographical distribution of ground stations on the surface. Based on the non-uniform distribution of different ground stations, a smaller number of satellites can be used to cover all the ground stations. Moreover, the distribution of satellites in space is dynamic. The position of satellites in space varies with the orbit. If the geographical distribution of ground stations is known. By adjusting the orbital parameter, the less satellites could cover more ground stations.

Currently, the satellite-to-ground quantum communication approach involves a single satellite transmitting to two ground stations. The challenge lies in the spatial separation of ground stations, often too distant for a single satellite to simultaneously cover. Traditionally, this necessitates a waiting period for the satellite to reposition itself to communicate with the second station. It needs to wait for the satellite to transmit a photon to one ground station and another pair of photons to the other ground station when the ground station moves over the other target ground station, a process that, depending on the Low Earth Orbit (LEO) satellite's altitude, can range from a few minutes to an hour, introducing significant delays in photon pair creation for quantum key distribution. To overcome this latency, we propose a network of 2-3 satellites orbiting along the equator, functioning as quantum relays. These satellites, equipped with quantum communication technology, including entangled photon generators and quantum repeaters\cite{dur1999quantum}, enable a seamless transfer of quantum information. The orbital mechanics are meticulously designed so that at least one satellite is always optimally positioned to interact with any given ground station, ensuring continuous coverage and significantly reducing transmission delays. In addition, we assume that the ground station can be synchronized with the satellite network for precise timing, which is a key factor in maintaining the efficiency of quantum key distribution.

The current focus of optimization studies in quantum satellite communication is on enhancing the instantaneous efficiency of satellite-to-ground transmissions. However, the position of satellites in space is continually shifting due to their orbital movements. This leads to a limitation in optimizing instantaneous photon transmission efficiency: the most efficient communication at any given moment does not guarantee optimal total photon transmission over a period. For a satellite in a circular orbit, while the distance from the Earth's surface remains relatively constant throughout its orbit, the changing elevation angle between the ground station and the satellite causes variations in the distance between them, and consequently, the communication efficiency fluctuates. Therefore, in our experiments, we keep the satellite's orbital period constant and assume that a coordination algorithm among the satellites is in place to prevent collisions\cite{li2023networking}. For the same satellite, we measure the total number of photons transmitted to a ground station over one orbital period, using this total photon count as the standard to evaluate the effectiveness of our proposed solution.

In our current approach, while maintaining the satellite's rotation period, we utilize the Molniya orbit to enhance communication efficiency. The Molniya orbit, characterized by its high eccentricity, significantly alters the satellite's distance from the Earth. This change serves a dual purpose: it expands the communication coverage and prolongs the satellite's presence over a particular hemisphere, thereby improving its ability to maintain continuous communication within that region. The increased distance achieved through the Molniya orbit is advantageous in terms of broader coverage, ensuring a more extensive and reliable communication network. Conversely, the Molniya orbit can also be employed to bring the satellite closer to the Earth, particularly beneficial for interactions with Earth relay stations. This closer proximity can result in more efficient quantum satellite communication, as it reduces the signal degradation that typically occurs over longer distances. The optimization of such an orbit is critical in this context, necessitating a careful balance between maximizing coverage and minimizing transmission losses. In our work, we strive to find the optimal satellite constellation parameters to maximize the communication efficiency of satellite networks

Here is the contribution of this paper:
\begin{itemize}
  \item \textbf{Innovative Quantum Relay Network:} We introduce a novel network design involving 2-3 satellites in equatorial orbit, functioning as quantum relays. This design significantly reduces latency in satellite-to-ground quantum communication and ensures continuous coverage with optimal positioning for communication with ground stations.

  \item \textbf{Optimization of Photon Transmission Efficiency:} Our research shifts the focus from instantaneous transmission efficiency to optimizing the total photon transmission over an entire orbital period. This approach offers a more comprehensive evaluation of the communication efficiency of quantum satellite networks.

  \item \textbf{Strategic Use of Molniya Orbit:} We utilize the Molniya orbit for expanding communication coverage and improving the operational presence of satellites over targeted hemispheres. This strategic orbital choice enhances continuous communication capabilities and addresses the limitations of traditional satellite orbits.
\end{itemize}

\section{Satellite Coverage and Communication Modeling}
This section is dedicated to detailing the methodology and models used in optimizing satellite coverage and communication efficiency. It consists of three subsections, each addressing a specific aspect of the satellite-ground station communication network.

\subsection{City Clustering for Optimal Satellite Coverage}
\label{subsec:city_clustering}
This subsection focuses on the grouping of fifteen selected cities to determine the optimal positioning of satellites. The aim is to cluster these cities into groups based on minimum earth surface coverage requirements, thereby facilitating better satellite arrangement and ensuring comprehensive coverage.

The primary objective is to group cities such that each cluster has cities not more than 400 km apart from each other. This constraint is critical for ensuring that a single satellite can cover all cities within a cluster.
To achieve this, we employ the Haversine formula for calculating the great-circle distance between each pair of cities, given their latitude and longitude. Subsequently, we apply the DBSCAN clustering algorithm, which is particularly suited for spatial data clustering with a specific distance threshold.

The Haversine formula calculates the distance between two points on a sphere given their longitudes and latitudes. It is given by:

\[ d = 2r \arcsin\left(\sqrt{\sin^2\left(\frac{\Delta\phi}{2}\right) + \cos(\phi_1)\cos(\phi_2)\sin^2\left(\frac{\Delta\lambda}{2}\right)}\right) \]

where \( \Delta\phi \) and \( \Delta\lambda \) are the differences in latitude and longitude, \( r \) is the Earth's radius, and \( d \) is the distance between the two points.

DBSCAN (Density-Based Spatial Clustering of Applications with Noise) forms clusters based on the density of data points. Points that are closely packed together are grouped into the same cluster. The algorithm parameters include:

\begin{itemize}
    \item \textit{eps}: The maximum distance between two samples for one to be considered as in the neighborhood of the other (set to 400 km in our case).
    \item \textit{min\_samples}: The number of samples in a neighborhood for a point to be considered as a core point.
\end{itemize}

Below is the pseudocode for the clustering process:

\begin{algorithm}\label{algorithm 1}
\caption{City Clustering Algorithm}
\begin{algorithmic}[1]
\Function{DBSCANClustering}{$distanceMatrix$, $eps$, $minSamples$}
    \State Initialize clusters as empty list
    \State Label all points as 'unvisited'
    \For{each point $P$ in $distanceMatrix$}
        \If{$P$ is labeled 'unvisited'}
            \State Label $P$ as 'visited'
            \State $NeighborPts \gets$ \Call{RegionQuery}{$P$, $eps$}
            \If{size of $NeighborPts$ $<$ $minSamples$}
                \State Label $P$ as 'noise'
            \Else
                \State $newCluster \gets$ \Call{ExpandCluster}{$P$, $NeighborPts$, $eps$, $minSamples$}
                \State Add $newCluster$ to clusters
            \EndIf
        \EndIf
    \EndFor
    \State \Return clusters
\EndFunction

\Function{ClusterCities}{$cities$, $eps$, $minSamples$}
    \State $distanceMatrix \gets$ \Call{CalculateDistanceMatrix}{$cities$}
    \State $clusters \gets$ \Call{DBSCANClustering}{$distanceMatrix$, $eps$, $minSamples$}
    \State \Return $clusters$
\EndFunction
\end{algorithmic}
\end{algorithm}

Upon applying this methodology, we successfully grouped the fifteen cities into clusters where each city is within a 250 km radius of every other city as 12 clusters. This clustering enables us to strategically position satellites to cover each cluster effectively, thereby optimizing our satellite network for comprehensive global coverage.

\subsection{Modeling Satellite-Ground Station Communication Links}
\label{subsec:satellite_ground_station_communication}

In this subsection, we present our approach for establishing and modeling the communication links between a single satellite and various ground stations. Our focus is on calculating the orbital periods of satellites, determining the slant range distances, and estimating the key generation rates based on the elevation angles and distances involved.

\paragraph{Orbital Period Calculation:}
The orbital period of a satellite is critical for understanding its operational dynamics. We calculate it based on the satellite's altitude above Earth's surface using the following formula:

\[ T = 2\pi \sqrt{\frac{(R + h)^3}{GM}} \]

where \( R \) is the Earth's radius, \( h \) is the satellite's altitude, \( G \) is the gravitational constant, and \( M \) is the Earth's mass. 

\paragraph{Slant Range Distance:}
The slant range distance, which is the direct line-of-sight distance between the satellite and a ground station, is calculated as follows:

\[ d = \sqrt{(R + h)^2 - (R \cos(\theta))^2} - R \sin(\theta) \]

where \( \theta \) is the elevation angle, \( R \) is the Earth's radius, and \( h \) is the satellite's altitude.

\paragraph{Quantum Transmission Efficiency}
The transmission efficiency is affected by the channel loss. Signal-loss ratio and the system error.\cite{dequal2021feasibility}. In this paper, we will use the existing satellite-to-ground communication link efficiency.\cite{liao2017satellite}. According Liao\cite{liao2017satellite}'s work, the satellite QKD link efficiencies

\subsection{Channel Loss and Quantum Generation Rate Models}
Here, we describe the models used for calculating channel loss in satellite-to-ground communication and the quantum generation rate. This subsection covers the theoretical framework and practical considerations in modeling the efficiency and reliability of quantum communication channels. The transmission efficiency is affected by the channel loss. Signal-loss ratio and the system error.\cite{dequal2021feasibility}. Liao et al\cite{liao2017satellite}.'s work indicates that the communication efficiency of satellites varies linearly with altitude, with recorded efficiencies of 12 Kbits at 645 km and 1 Kbit at 1200 km. Utilizing this data, we can fit a model to predict the QKD link efficiency (in dB) for any given distance between a satellite and a ground station. This relationship is modeled as:
\[ \text{Link Efficiency (LE)} = \text{slope\_efficiency} \times (\text{distance} - D) \]

where slope\_efficiency is calculated based on the known data points. \(D\) is the original data points in the Micius Satellite's experiment data.

To determine the quantum generation rate (\( T \)) at different altitudes, we combine the initial generation rate (\( T_0 \)) with the link efficiency (\( LE \)). The generation rate is modeled as:

\[ T = T_0 \times 10^{\frac{\text{LE}}{10}} \]

where \( T_0 \) is the baseline quantum key generation rate when the satellite distance is \(D\), and \( LE \) is the link efficiency calculated from the distance.

\subsection{Experimental Analysis}
\label{subsec:experimental_analysis}

In this subsection, we present the experimental findings from two key studies that investigate the relationship between quantum link efficiency and two factors: elevation angle and slant range. The altitude of the satellite was held constant at 400 km for both experiments.

In the first experiment, the elevation angle was varied from 5 to 90 degrees. The resulting key generation rates at these angles were calculated and plotted.

\begin{figure}[H]
\centering
\includegraphics[width=0.45\textwidth]{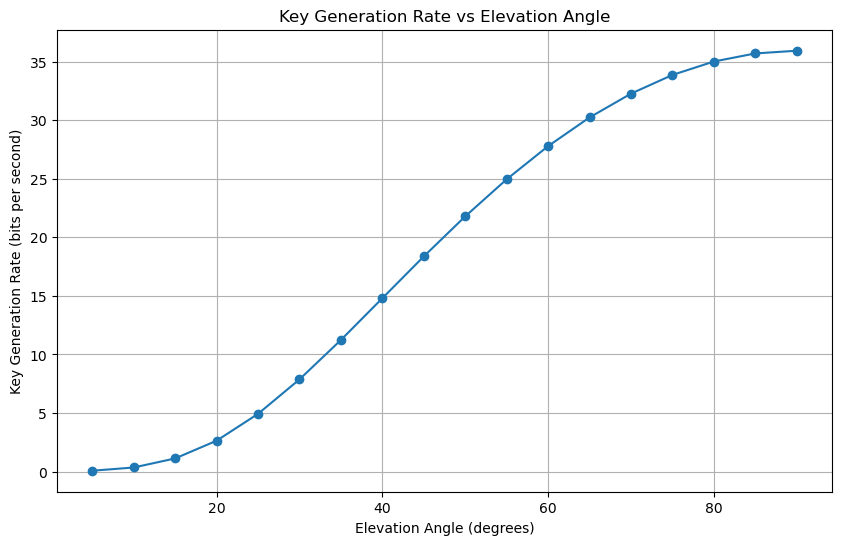}
\caption{Key Generation Rate vs Elevation Angle}
\label{fig:key_gen_rate_vs_elevation_angle}
\end{figure}

As depicted in Figure \ref{fig:key_gen_rate_vs_elevation_angle}, the key generation rate increases with the elevation angle, indicating an improved quantum link efficiency at higher angles. Initially, at lower elevation angles, the key generation rate increases slowly. As the elevation angle continues to rise, the rate of increase becomes more pronounced. However, as the angle nears 90 degrees, the efficiency gains in the link are not as significant. This pattern of efficiency improvement can be attributed to the increase in the atmospheric thickness within the link as the elevation angle from the satellite to the ground station increases. We also observed that when the elevation angle between the satellite and the ground station is less than 20 degrees, the key generation rate is relatively low. Therefore, we define the effective communication range for the satellite in terms of elevation angle as from 20 to 160 degrees.

The second experiment focused on the relationship between the key generation rate and the slant range, which is the direct distance from the satellite to the ground station. The slant range was computed for the fixed satellite altitude at various elevation angles.

\begin{figure}[H]
\centering
\includegraphics[width=0.45\textwidth]{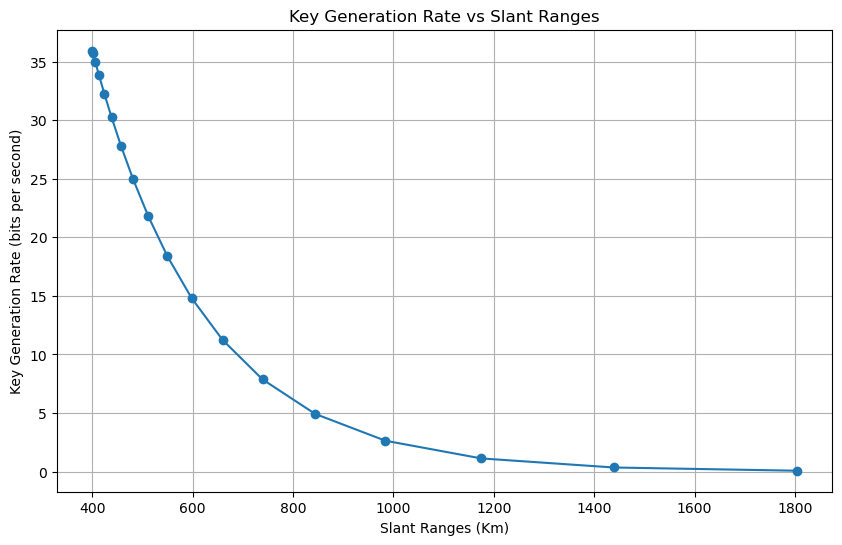}
\caption{Key Generation Rate vs Slant Range}
\label{fig:key_gen_rate_vs_slant_range}
\end{figure}

Figure \ref{fig:key_gen_rate_vs_slant_range} shows a clear inverse relationship between the key generation rate and the slant range. As the distance increases, the efficiency of the quantum link decreases, which can be attributed to factors such as beam divergence and atmospheric absorption.

\subsection{Total Photon Transmission Analysis Over One Orbital Period}
\label{subsec:photon_transmission_analysis}

In the concluding part of our experimental analysis, we evaluate the total number of photons that a satellite can successfully transmit to a ground station in one orbital period. Assuming a circular orbit, where the satellite invariably passes directly above the ground station, we identify an instance where the elevation angle reaches 90 degrees.

Given the previous simulations, we have established that the effective elevation angles for communication range from 20 to 160 degrees. By utilizing the satellite's altitude, we can calculate its angular velocity, denoted by \( \omega \), and subsequently determine the effective operational time window, \( t_1 \) to \( t_2 \), where the satellite is positioned within the effective elevation angles above the ground station.

The instantaneous transmission rate, \( T(t) \), depends on the time-variable distance between the satellite and the ground station, \( D(t) \), and is given by:
\begin{equation}
T(t) = T_0 \times 10^{\frac{\text{s} \cdot D(t) - D_0}{10}}
\end{equation}
where \( t \) represents the current time, \( D(t) \) the current distance to the ground station, and \( D_0 \) a reference distance obtained from fitted data of known quantum satellite systems.

To derive the total number of transmitted photon pairs over one effective period, we integrate \( T(t) \) from \( t_1 \) to \( t_2 \):
\begin{equation}
\text{Total Photon Pairs} = \int_{t_1}^{t_2} T(t) \, dt
\end{equation}

We perform this analysis for altitudes ranging from 400 to 1200 km to observe the relationship between altitude and photon transmission. The results are depicted in the graph below.

\begin{figure}[H]
\centering
\includegraphics[width=0.45\textwidth]{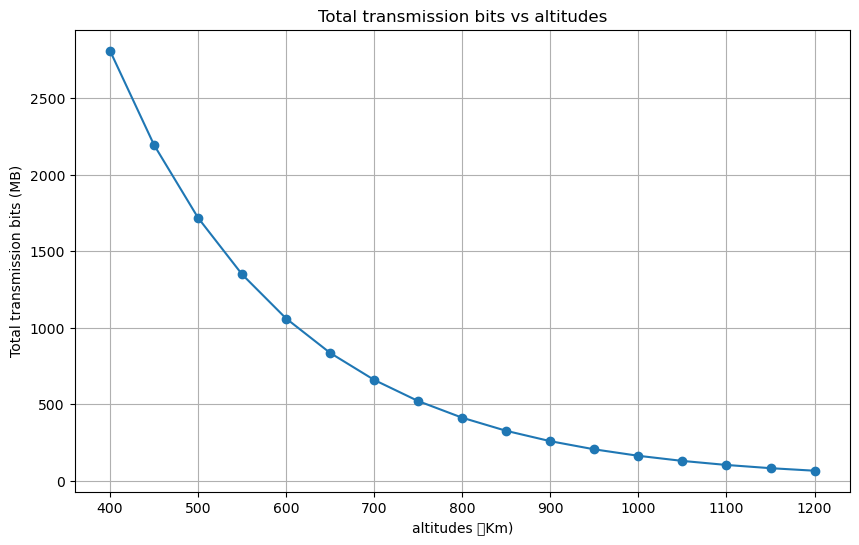}
\caption{Total Transmission Bits vs Altitudes}
\label{fig:total_transmission_bits_vs_altitudes}
\end{figure}

As shown in Figure \ref{fig:total_transmission_bits_vs_altitudes}, there is a notable decrease in total transmitted bits with increasing altitude. This trend highlights the impact of altitude on transmission efficiency, suggesting an optimal range for satellite operation to maximize photon transmission.

This analysis forms a critical component of our research, providing vital data to optimize satellite-based QKD systems, balancing altitude against transmission capabilities to enhance the security and reliability of quantum communication channels.

\section{Quantum Satellite Network Optimization}
\label{sec:quantum_satellite_network_optimization}

In our comprehensive analysis, we have delineated the coverage scope of an individual orbital satellite communication link. This encompasses the integration of satellite orbits predicated upon ground station positioning and a rigorous substantiation of communication efficiencies. However, achieving comprehensive coverage across various ground stations necessitates the intersection of multiple satellite orbits. Moreover, facilitating intersatellite communication across different orbits introduces inherent latency, often spanning several minutes or, in certain cases, resulting in a lack of orbital intersection.

To circumvent these challenges, our proposal integrates quantum relay satellites into the network. These satellites act as conduits, connecting quantum satellites on disparate orbits to ensure a cohesive network encompassing all ground stations within the service area. Quantum communication satellites are optimized for high-efficiency links, necessitating their proximity to the ground stations they serve. On the other hand, quantum relay satellites prioritize extensive coverage and maximized operational timeframes, suggesting a deployment strategy that places them at higher altitudes from their orbital paths.

Employing Molniya orbits for the relay satellites enhances the operational duty cycle within a single period, guaranteeing substantial coverage over the targeted ground stations. In this role, quantum relay satellites are solely tasked with photon transmission between satellites. This transmission, unfettered by atmospheric constraints, experiences significantly lower photon loss compared to ground-based links.

\section{Future Work}
As modern communication systems advance, wireless networks \cite{wire1,wire3,10017581, 9523755,9340574,10.1145/3387514.3405861,9141221,9120764,10.1145/3356250.3360046,8737525,8694952,10.1145/3274783.3274846,10.1145/3210240.3210346,8486349,8117550,8057109,https://doi.org/10.1155/2017/5156164,10189210} and secure communication \cite{wire2, 10125074,285483,10.1145/3395351.3399367} face increasing challenges in reliability and data protection. AI-driven methods and system optimization \cite{10.1145/3460120.3484766,9709070,9444204,ning2021benchmarkingmachinelearningfast,8832180,8556807,8422243,chandrasekaran2022computervisionbasedparking,iqbal2021machinelearningartificialintelligence,pan2020endogenous} are becoming essential tools to address these issues by enhancing performance and mitigating potential threats. In this context, QKD emerges as a revolutionary cryptographic approach, leveraging quantum mechanics to achieve unparalleled security. By utilizing principles such as the no-cloning theorem and the Heisenberg uncertainty principle, QKD ensures robust defense against eavesdropping and tampering, positioning quantum communication at the forefront of future-proof secure technologies. In the future, we will focus on leveraging AI-driven optimization and system-level coordination to enhance satellite-based quantum communication networks and integrating wireless networking techniques with QKD to improve the scalability and flexibility of satellite-ground communication. In this case, future research can bridge the gap between theoretical advancements and practical deployment of global-scale quantum-secure networks.

\section{Conclusion}
The findings of this study contribute significantly to the field of satellite-based quantum communication. We have successfully demonstrated a methodology for clustering cities to optimize satellite coverage and outlined a model for satellite-to-ground station communication links. Our experimental analysis confirms the dependency of quantum link efficiency on elevation angle and slant range, guiding the strategic positioning of satellites. By incorporating quantum relay satellites and employing Molniya orbits, we propose a solution to overcome the latency issues inherent in intersatellite communication across different orbits. This network topology ensures extensive coverage and high-efficiency links, essential for the practical implementation of a global quantum communication network. Our research lays the groundwork for future developments in satellite QKD systems, paving the way for secure and reliable quantum communications across the globe.

\bibliographystyle{unsrt}  
\bibliography{Citation, zhu}  

\end{document}